\documentclass[english,10pt,aps,prd,a4paper,preprintnumbers,floatfix,nofootinbib,showpacs,superscriptaddress, notitlepage, twocolumn]{revtex4-1}
\pdfoutput=1

\usepackage{graphicx}
\usepackage{rotate}
\usepackage[utf8]{inputenc}
\usepackage{amsmath}
\usepackage{amssymb}
\usepackage{float}
\usepackage{comment}
\usepackage{soul}
\usepackage[usenames,dvipsnames]{color}
\usepackage[colorinlistoftodos]{todonotes}
\usepackage{amsmath}
\usepackage{amssymb}
\usepackage[colorlinks=true,citecolor=darkred,urlcolor=darkred, pdfborder={0 0 0}]{hyperref}
\newcommand {\ignore}[1]{}
\definecolor{darkred}{rgb}{0.6,0,0}

\def\tt1{$\mathrm{SU(3) \otimes SU(3)_L \otimes U(1)}$ }
\def\3311{$\mathrm{SU(3) \otimes SU(3)_L \otimes U(1)_X \otimes U(1)_{N}}$ }
\def\gsim{\raise0.3ex\hbox{$\;>$\kern-0.75em\raise-1.1ex\hbox{$\sim\;$}}}
\def\lsim{\raise0.3ex\hbox{$\;<$\kern-0.75em\raise-1.1ex\hbox{$\sim\;$}}}

\definecolor{mightnightblue}{RGB}{25,25,112}
\definecolor{brown}{rgb}{0.59, 0.29, 0.0}

\def\21{$\mathrm{SU(2)_L \otimes U(1)_Y}$}

\newcommand{\AddrAHEP}{%
  AHEP Group, Institut de F\'{i}sica Corpuscular --
  CSIC/Universitat de Val\`{e}ncia, Parc Cient\'ific de Paterna.\\
 C/ Catedr\'atico Jos\'e Beltr\'an, 2 E-46980 Paterna (Valencia) - SPAIN}

\begin{document}
\bibliographystyle{unsrt}   
\title{Predicting neutrino oscillations with ``bi-large'' lepton mixing matrices }
\author{Peng Chen}\email{pche@mail.ustc.edu.cn}
\affiliation{College of Information Science and Engineering,Ocean University of China, Qingdao 266100, China}
\author{Gui-Jun Ding}\email{dinggj@ustc.edu.cn}
\affiliation{Interdisciplinary Center for Theoretical Study and Department of Modern Physics, \\
University of Science and Technology of China, Hefei, Anhui 230026, China}
\author{Rahul Srivastava}\email{rahulsri@ific.uv.es}
\affiliation{\AddrAHEP}
\author{Jos\'{e} W. F. Valle}\email{valle@ific.uv.es}
\affiliation{\AddrAHEP}

\begin{abstract}
\vspace{1cm}

We propose two schemes for the lepton mixing matrix $U = U_l^\dagger U_\nu$, where $U = U_l$ refers to the charged sector, and $U_\nu$ denotes the neutrino diagonalization matrix. We assume $U_\nu$ to be CP conserving and its three angles to be connected with the Cabibbo angle in a simple manner.
CP violation arises solely from the $U_l$, assumed to have the CKM form, $U_l\simeq V_{\rm CKM}$, suggested by unification. Oscillation parameters depend on a single parameter, leading to narrow ranges for the ``solar'' and ``accelerator'' angles $\theta_{12}$ and $\theta_{23}$, as well as for the CP phase, predicted as $\delta_{\rm CP}\sim 1.3\pi$.

\end{abstract}
\preprint{USTC-ICTS-19-04}
\maketitle

\section{Introduction}

After decades of hard work, the origin of flavor mixing and CP violation remains one of the most important challenges in particle physics. Understanding the flavor problem would help us to get a glimpse on physics beyond the standard model.
Several approaches have been pursued to find an adequate and predictive description of lepton mixing.
Using flavor symmetries from first principles~\cite{Ishimori:2010au} one can obtain ``top down'' restrictions on neutrino mixing within fundamental theories of neutrino mass~\cite{Babu:2002dz,Altarelli:2010gt,Chen:2015jta,Morisi:2012fg,King:2013eh}.
Alternatively, one may make educated phenomenological guesses as to what the pattern of lepton mixing should look like.
Specially influential were the ideas of mu-tau symmetry and the Tri-Bimaximal (TBM) lepton mixing ansatz proposed by Harrsion, Perkins and Scott~\cite{Harrison:2002et,Harrison:2002er,harrison:2002kp}.
The latter predicts the three mixing angles as $\sin^2\theta_{23}=1/2$, $\sin^2\theta_{12}=1/3$, $\sin^2\theta_{13}=0$, while the Dirac CP phase vanishes.
However, the precise measurements of the reactor angle $\theta_{13}\sim8.5^\circ$ in Daya Bay~\cite{An:2016ses}, RENO~\cite{Pac:2018scx} and Double Chooz~\cite{Abe:2014bwa} now exclude TBM as a realistic lepton mixing pattern.
The discrepancy between experiment and the prediction of TBM led people to pursue new lepton mixing structures. One method is to modify the TBM pattern based on flavor or CP symmetries such as in ref.~\cite{Chen:2018eou,Chen:2018zbq}.

A more phenomenological approach is to explore new neutrino mixing patterns~\cite{Minakata:2004xt,Albright:2010ap,Plentinger:2005kx}.
Recently a ``bi-large'' mixing scheme has been proposed in Ref.~\cite{Boucenna:2012xb} assuming that $\sin\theta_{13}\simeq\lambda$ where $\lambda$ is the Cabibbo angle.
A generalization of this pattern was proposed in Ref.~\cite{Roy:2014nua}, taking the Cabibbo angle as a universal seed for quark and lepton mixing. Such schemes may emerge from Grand Unified Theories (GUTs) and flavor symmetry~\cite{Ding:2012wh}.
The good features of such bi-large mixing patterns deserve further investigation.

In this paper we will propose two ``bi-large'' lepton mixing schemes and investigate their phenomenological implications. For definiteness, we assume normal ordered neutrino masses throughout this paper, since inverted ordering is disfavored at more than $3\sigma$~\cite{deSalas:2017kay}. As in Ref.~\cite{Roy:2014nua}, we assume that the charged lepton diagonalization matrix is CKM-like, given in terms of the Wolfenstein parameters $\lambda$ and $A$, whose values we take from the PDG as $\lambda=0.22453$ and $A=0.836$~\cite{Tanabashi:2018oca}. In our ansatz the three mixing angles characterizing the neutrino diagonalization matrix are related with $\lambda$ in a very simple manner.
We obtain tight predictions for the physical lepton mixing angles and the CP phase. These are contrasted with current experiments and used to make projections for upcoming long baseline oscillation experiments.


\section{Bi-large: pattern I}
\label{sec:bi-large:-pattern-1}


In this section we propose our first ``bi-large'' lepton mixing pattern. Within the standard parameterization the three angles of the neutrino diagonalization matrix are assumed to be given as
\begin{eqnarray}
\label{eq:bl1}\sin\theta^\nu_{23}=1-\lambda\,,\quad\sin\theta^\nu_{12}=2\lambda\,,\quad\sin\theta^\nu_{13}=\lambda\,,
\end{eqnarray}
and the Dirac CP phase is taken as $\delta^\nu_{CP}=\pi$~\footnote{The Majorana phases are taken to be zero, hence our ansatz is just for oscillation physics, with hardly any predictivity for neutrinoless double beta decay experiments.}.
In this case the neutrino diagonalization matrix can be approximated as
\begin{eqnarray}
&&\label{eq:unu1}U_{\nu}\simeq\left(\begin{matrix}
1-\frac{5\lambda^2}{2}&2\lambda&-\lambda\cr
\lambda-2\sqrt{2}\lambda^{\frac{3}{2}}&\sqrt{2\lambda}-\frac{\lambda^{\frac{3}{2}}}{2\sqrt{2}}&1-\lambda-\frac{\lambda^2}{2}\cr
2\lambda+\sqrt{2}\lambda^{\frac{3}{2}}&-1+\lambda&\sqrt{2\lambda}-\frac{\lambda^{\frac{3}{2}}}{2\sqrt{2}}\cr
\end{matrix}\right)~.
\end{eqnarray}
If the charged leptons are taken diagonal then this will imply that the leptonic mixing parameters are the same as eq.~\eqref{eq:bl1}: $\sin^2\theta_{23}=\sin^2\theta^\nu_{23}\simeq0.601$, $\sin^2\theta_{12}=\sin^2\theta^\nu_{12}\simeq0.202$ and $\sin^2\theta_{13}=\sin^2\theta^\nu_{13}\simeq0.0504$, and lie outside the $3\sigma$ experimental range~\cite{deSalas:2017kay}. However, corrections are expected from the charged lepton diagonalization matrix.
Following Ref.~\cite{Roy:2014nua} we assume that the bi-large pattern arises from the simplest SO(10) model where the charged and the down-type quarks have roughly the same mass. Then the lepton diagonalization matrix is naturally of the CKM type
\begin{eqnarray}
\label{eq:ul1}
\begin{aligned}
U_{l}&= R_{23}(\theta^{CKM}_{23})\,\Phi\,R_{12}(\theta^{CKM}_{12})\,\Phi^\dagger\\
&\simeq\left(\begin{array}{ccc}
1-\frac{\lambda^2}{2}&\lambda e^{-i\phi}&0\\
-\lambda e^{i\phi}&1-\frac{\lambda^2}{2}&A\lambda^2\\
A\lambda^3e^{i\phi}&-A\lambda^2&1
\end{array}\right)\,,
\end{aligned}
\end{eqnarray}
where $\sin\theta^{CKM}_{23}=A\lambda^2$ and $\sin\theta^{CKM}_{12}=\lambda$, where $\lambda$ and $A$ are the Wolfenstein parameters, $R_{ij}$ is the $i$-$j$ real rotation matrix, and $\Phi=\text{diag}(e^{-i\phi/2},e^{i\phi/2},1)$ where $\phi$ is a free phase. For convenience we set $\phi\in(-\pi,\pi]$ throughout this paper. The elements of the lepton mixing matrix $U = U_l^\dagger U_{\nu}$ are all given in terms of just one free parameter $\phi$,
leading to a very high degree of predictivity.

To leading order the three mixing angles and the Jarlskog invariant $J_{CP}$ are given by
\begin{eqnarray}
\begin{aligned}
\sin^2\theta_{13} &\simeq 4\lambda^2 (1-\lambda)\cos^2\frac{\phi}{2}\,,\\
\sin^2\theta_{12} &\simeq 2\lambda^2\big(2-2\sqrt{2\lambda}\cos\phi+\lambda\big)\,,\\
\sin^2\theta_{23} &\simeq (1-\lambda)^2-2\sqrt{2}A\lambda^\frac{5}{2}-2\lambda^3(1+2\cos\phi)\,,\\
J_{CP}&\simeq-2\sqrt{2}\lambda^{\frac{5}{2}}\sin\phi\,.
\end{aligned}
\end{eqnarray}
The fact that the above parameters depend on just one free parameter $\phi$, leads to strong correlations.
\begin{figure}[h!]
\begin{center}
\begin{tabular}{c}
\includegraphics[width=.8\linewidth]{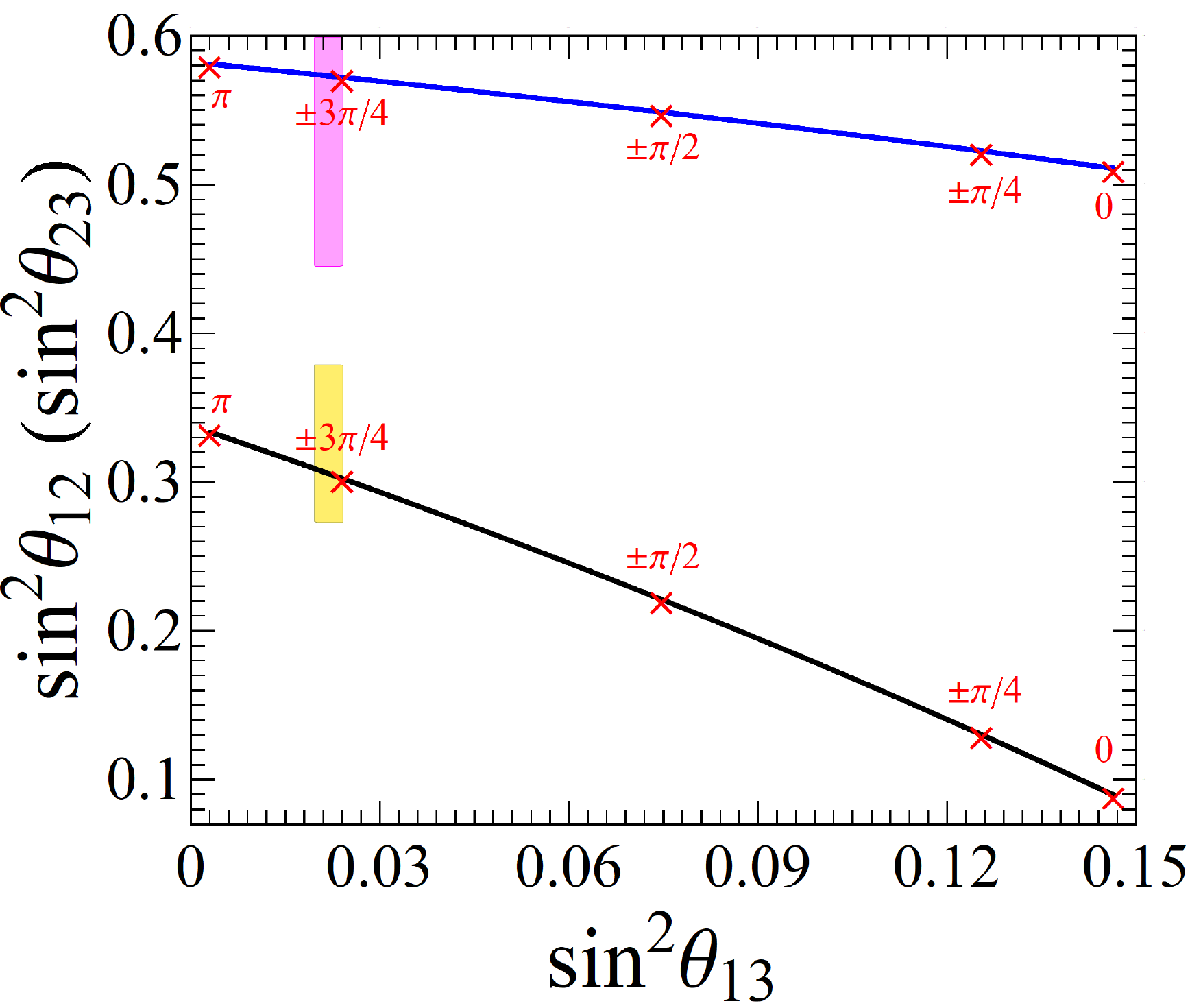}
\end{tabular}
\caption{\label{fig:angleI}
Predicted correlations involving the three mixing angles in pattern I.
``Solar'' and ``accelerator'' angles $\sin^2\theta_{12}$ (lower, black line) and $\sin^2\theta_{23}$ (upper, blue line)
correlate with the ``reactor'' mixing parameter $\sin^2\theta_{13}$.
The yellow and the magenta boxes represent the current 3$\sigma$ ranges of the mixing angles~\cite{deSalas:2017kay}. The cross symbols correspond to $\phi=0, \pi, \pm\pi/4, \pm\pi/2, \pm3\pi/4$, respectively.}
\end{center}
\end{figure}
\begin{figure}[h!]
\begin{center}
\begin{tabular}{c}
\includegraphics[width=0.8\linewidth]{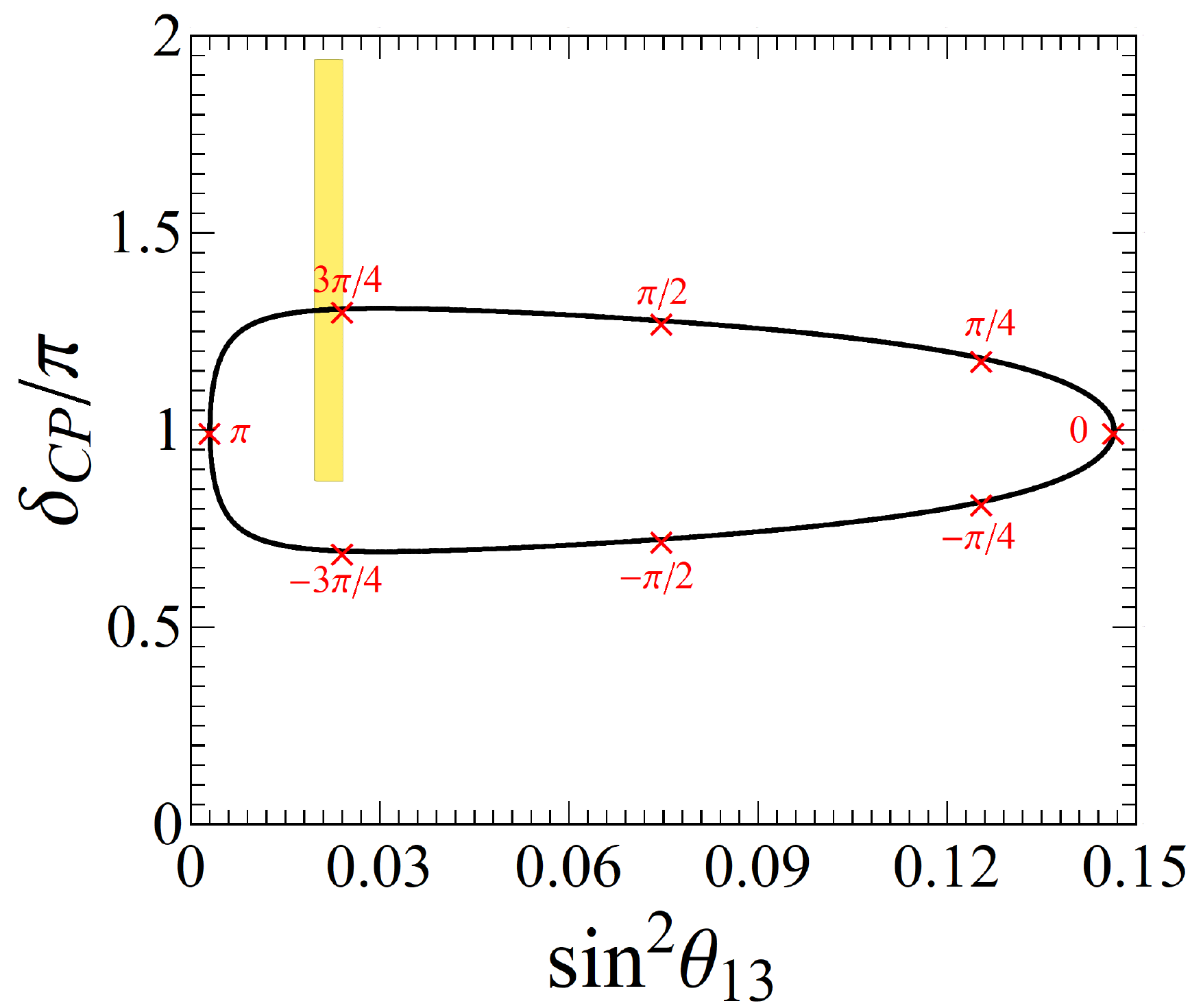}
\end{tabular}
\caption{\label{fig:dcpI}
Predicted correlation between the Dirac phase $\delta_{CP}$ and the ``reactor'' mixing parameter $\sin^2\theta_{13}$ in pattern I.
The yellow box is the current 3$\sigma$ range from the global fit~\cite{deSalas:2017kay}. The crosses correspond to $\phi=0, \pi,  \pm\pi/4, \pm\pi/2, \pm3\pi/4$, respectively.}
\end{center}
\end{figure}

The predictions for the oscillation parameters are shown in Fig.~\ref{fig:angleI} and Fig.~\ref{fig:dcpI}.
Requiring $\sin^2\theta_{13}$ to lie inside the allowed $3\sigma$ range implies that the value of $\phi/\pi$ should be inside the range of $[0.750,0.779]$. 
This severely restricts the allowed ranges for the $\theta_{12}$, $\theta_{23}$ and $\delta_{CP}$. The resulting ranges for the oscillation parameters become
\begin{eqnarray}
\begin{aligned}
0.0196\leq\sin^2\theta_{13}\leq0.0241\,,\\
0.302\leq\sin^2\theta_{12}\leq0.309\,,\\
0.572\leq\sin^2\theta_{23}\leq0.574\,,\\
1.303\leq\delta_{CP}/\pi\leq1.307\,.
\end{aligned}
\end{eqnarray}
One sees that, given $\theta_{13}$, we find that the resulting allowed ranges for the other mixing angles and CP violation phase are very narrow. The $\chi^2$ takes the minimum value $\chi^2_{\rm min}=2.394$ when $\phi=0.766\pi$, leading to the following values for the physical mixing parameters,
\begin{eqnarray}
\begin{aligned}
\sin^2\theta_{23}=0.573\,,\quad
\sin^2\theta_{13}=0.0216\,,\\
\sin^2\theta_{12}=0.306\,,\quad
\delta_{CP}=1.305\pi\,,
\end{aligned}
\end{eqnarray}
where $\sin^2\theta_{13}$, $\sin^2\theta_{12}$ and $\delta_{CP}$ are inside the $1\sigma$ range
while $\sin^2\theta_{23}$ is inside the $2\sigma$ range of~\cite{deSalas:2017kay}, hence fitting very well the experimental results.
One sees that the three mixing angles and the Dirac phase are in very good agreement with the current experimental values~\cite{deSalas:2017kay}.
It is also remarkable that, starting from a  CP conserving $U_\nu$ in eq.~\eqref{eq:unu1}, we obtain a CP violating phase that lies very close to the best fit value.

\section{Bi-large: pattern II}
\label{sec:bi-large:-pattern-2}

\begin{figure}[h!]
\begin{center}
\begin{tabular}{c}
\includegraphics[width=0.8\linewidth]{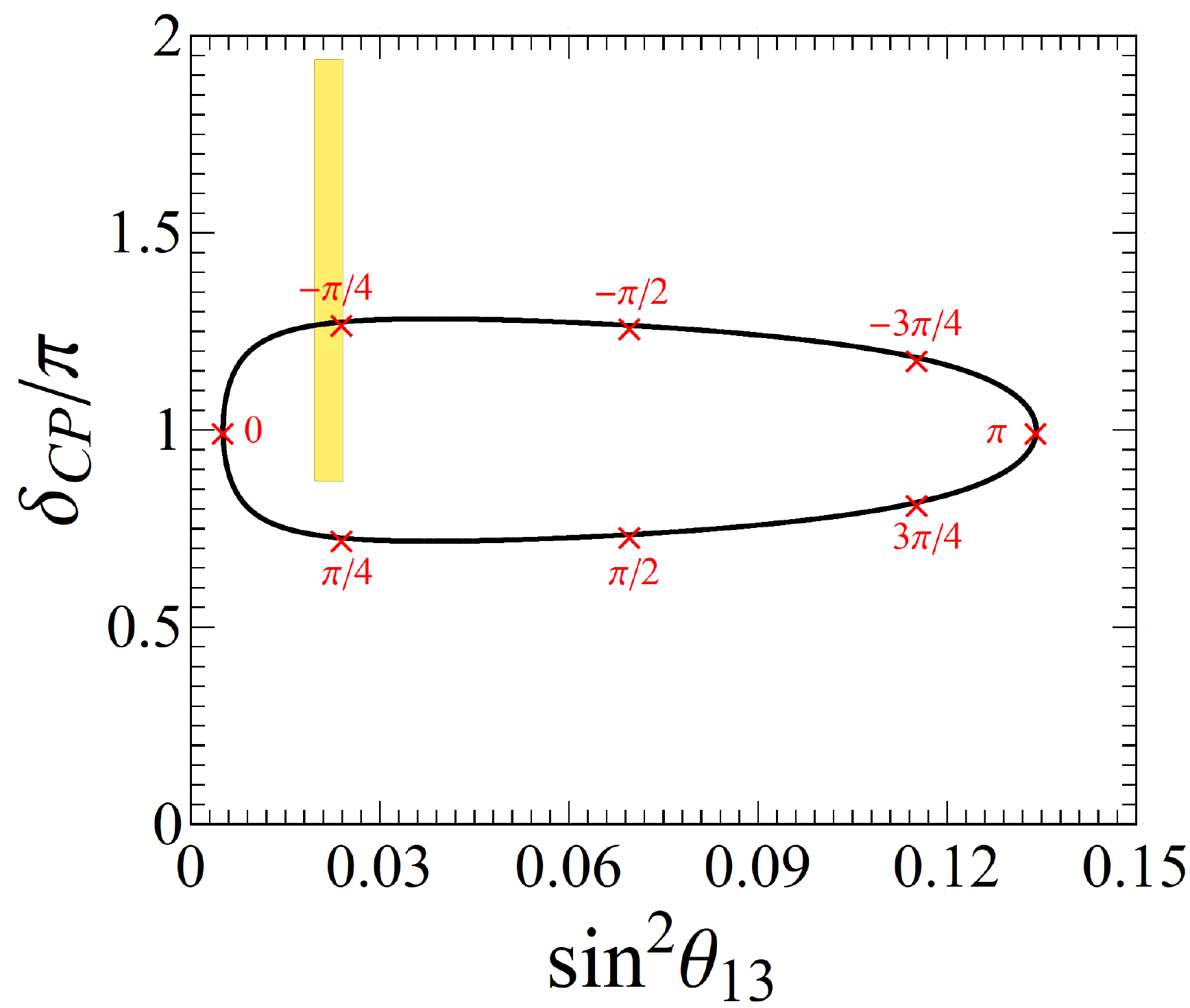}
\end{tabular}
\caption{\label{fig:ang_phase}
Predicted correlation between the Dirac phase $\delta_{CP}$ and the ``reactor'' mixing parameter $\sin^2\theta_{13}$ in pattern II.
The yellow box is the current 3$\sigma$ range from the global fit~\cite{deSalas:2017kay}. The crosses correspond to $\phi=0, \pi,  \pm\pi/4, \pm\pi/2, \pm3\pi/4$, respectively.}
\end{center}
\end{figure}

We now turn to our second example. Again we take the Dirac CP phase as $\delta^\nu_{CP}=\pi$ but now assume the neutrino mixing angles in the standard parameterization to be given by
\begin{eqnarray}
\label{eq:bl2}
\sin\theta^\nu_{13} = 1\lambda\,, \quad \sin\theta^\nu_{12} = 2\lambda\,, \quad \sin\theta^\nu_{23}=3\lambda\,.
\end{eqnarray}
To order $\lambda^2$ the neutrino mixing matrix of such ``1-2-3'' bi-large mixing pattern is written as
\begin{eqnarray}
U_{\nu}\simeq\left(\begin{array}{ccc}
1-\frac{5\lambda^2}{2} & 2\lambda & -\lambda \\
-2\lambda+3\lambda^2 & 1-\frac{13\lambda^2}{2} & 3\lambda \\
\lambda+6\lambda^2 & -3\lambda+2\lambda^2 & 1-5\lambda^2 \\
\end{array}\right)\,.
\end{eqnarray}
As theoretical motivation this time we consider the framework of SU(5) Grand Unified models. In the simplest SU(5) GUTs the lepton and down quark mass
matrices obey the relation $M_{e}\sim M_{d}^{T}$. As in the previous section,
this suggests us to adopt a CKM-type lepton diagonalization matrix
\begin{eqnarray}
\begin{aligned}
U_{l}&=\Phi^\dagger\,R_{12}^T(\theta^{CKM}_{12})\,\Phi\,R_{23}^T(\theta^{CKM}_{23})\\
&\simeq\left(\begin{array}{ccc}
1-\frac{\lambda^2}{2}&-\lambda e^{i\phi}&A\lambda^3 e^{i\phi}\\
\lambda e^{-i\phi} & 1-\frac{\lambda^2}{2} & -A\lambda^2\\
0&A\lambda^2&1
\end{array}\right)\,,
\end{aligned}
\end{eqnarray}
with $\phi\in(-\pi,\pi]$. Then to leading order, the mixing angles and
$J_{CP}$ obtained from the lepton mixing matrix $U = U_l^\dagger U_\nu$ are given by
\begin{eqnarray}
\begin{aligned}
\sin^2\theta_{13} &\simeq \lambda^2 - 6\lambda^3 \cos\phi\,,\\
\sin^2\theta_{12} &\simeq \lambda^2 \big(5+4\cos\phi\big)\,,\\
\sin^2\theta_{23} &\simeq 9\lambda^2+6\lambda^3(A+\cos\phi)\,,\\
J_{CP} &\simeq -3\lambda^3\sin \phi\,.
\end{aligned}
\end{eqnarray}
\begin{figure}[h!]
\begin{center}
\begin{tabular}{c}
\includegraphics[width=.8\linewidth]{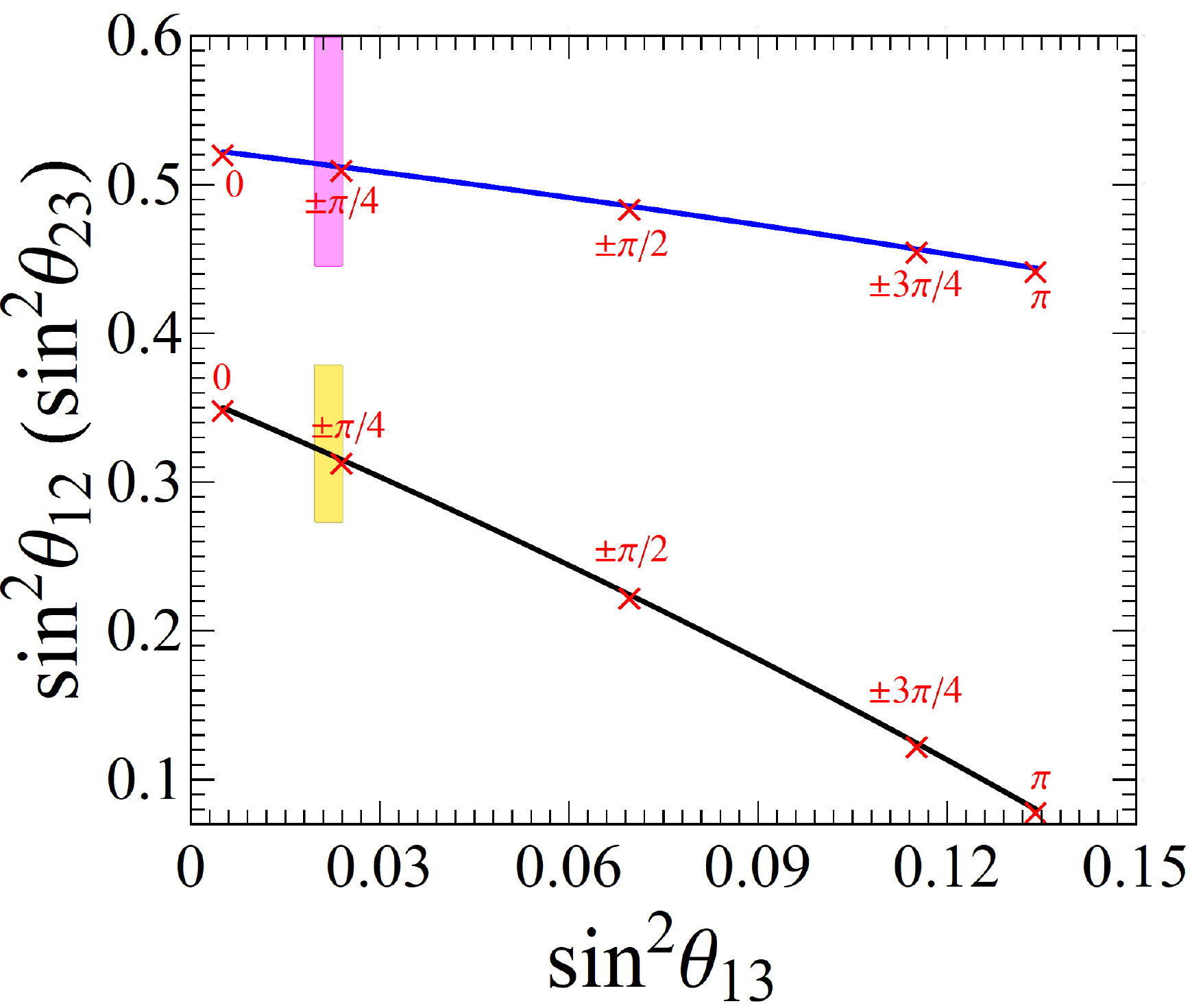}
\end{tabular}
\caption{\label{fig:mix_angles}
Predicted correlations involving the three mixing angles in pattern II. ``Solar'' and ``accelerator'' angles $\sin^2\theta_{12}$ (lower, black line) and $\sin^2\theta_{23}$ (upper, blue line) correlate with the ``reactor'' mixing parameter $\sin^2\theta_{13}$. The yellow and the magenta boxes represent the current 3$\sigma$ ranges of the mixing angles~\cite{deSalas:2017kay}. The cross symbols correspond to $\phi=0, \pi, \pm\pi/4, \pm\pi/2, \pm3\pi/4$, respectively.}
\end{center}
\end{figure}
As before, requiring $\sin^2\theta_{13}$ to lie in the allowed  3$\sigma$ range severely restricts the consistency ranges for the other oscillation parameters $\theta_{12}$, $\theta_{23}$ and $\delta_{CP}$, as follows
\begin{eqnarray}
\begin{aligned}
0.0196\leq\sin^2\theta_{13}\leq0.0241\,,\\
0.315\leq\sin^2\theta_{12}\leq0.323\,,\\
0.512\leq\sin^2\theta_{23}\leq0.514\,,\\
1.265\leq\delta_{CP}/\pi\leq1.274\,.
\end{aligned}
\end{eqnarray}
The $\chi^2$ takes the minimum value value $\chi^2_{\rm min}=2.954$
when $\phi=-0.232\pi$, and the mixing parameters are
\begin{eqnarray}
\begin{aligned}
\sin^2\theta_{23}=0.513\,,\quad\sin^2\theta_{13}=0.0216\,,\\
\sin^2\theta_{12}=0.320\,,\quad\delta_{CP}=1.270\pi\,.
\end{aligned}
\end{eqnarray}
One sees that $\sin^2\theta_{13}$, $\sin^2\theta_{12}$ and $\delta_{CP}$ are inside the $1\sigma$ range, while $\sin^2\theta_{23}$ is inside the $2\sigma$ range given by current global oscillation fits. The results are displayed in Figs.~\ref{fig:mix_angles} and \ref{fig:ang_phase}. As before, one sees that the predictions fit very well with the observed oscillation parameter values.


\section{long baseline oscillations}
\label{sec:osc}

\begin{figure}[h!]
\begin{center}
\begin{tabular}{cc}
\includegraphics[width=0.8\linewidth]{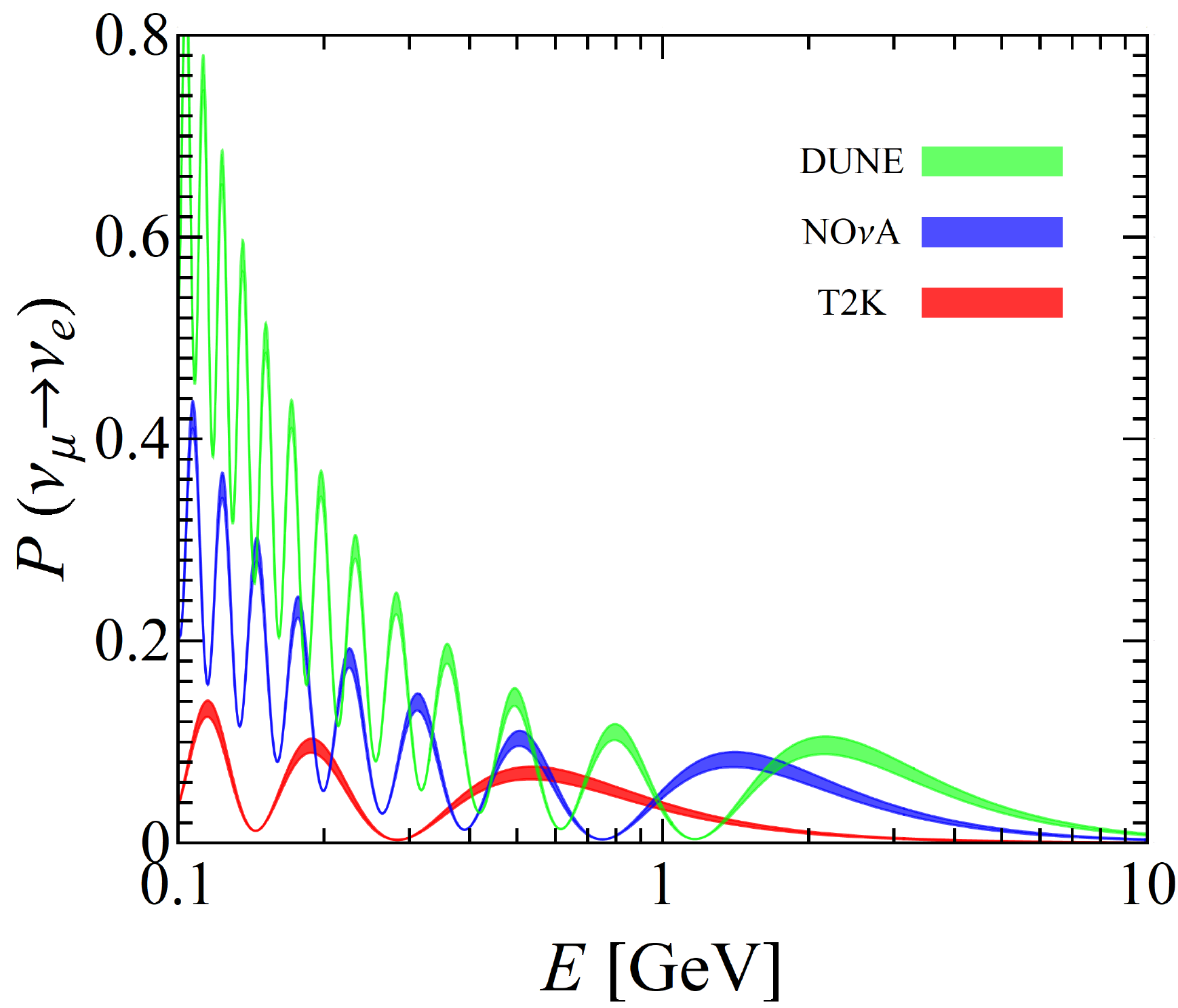}
\end{tabular}
\caption{\label{fig:pmueE}The $\nu_\mu \to \nu_e$ transition probability versus energy for pattern I for the T2K, NO$\nu$A and DUNE experiments. The mixing angles and Dirac CP phase are taken within the currently allowed 3$\sigma$ range. }
\end{center}
\end{figure}
\begin{figure}[h!]
\begin{center}
\begin{tabular}{cc}
\includegraphics[width=0.8\linewidth]{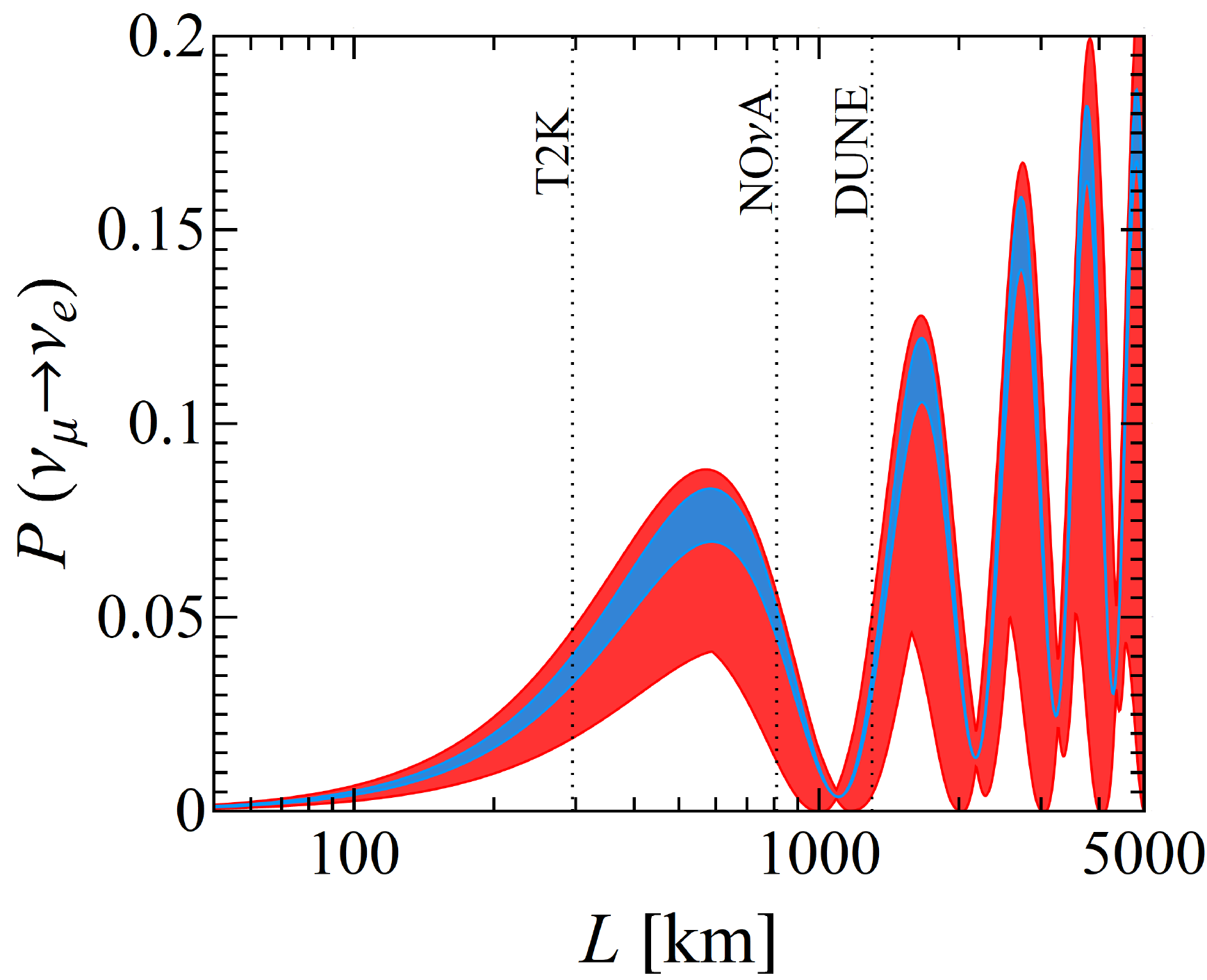}
\end{tabular}
\caption{\label{fig:pmueL}$\nu_\mu \to \nu_e$ transition probability versus distance taking the mixing angles and Dirac CP phase within the currently allowed 3$\sigma$ range. The broad band (red) refer to a generic scenario, whereas the thin band (blue) corresponds to the bi-large pattern I prediction. }
\end{center}
\end{figure}

The lepton mixing matrix in both cases discussed above only depends on one free parameter $\phi$.
As we saw, the one-parameter nature of both \textit{anzatze} leads to tight correlations amongst the oscillation parameters and predict very narrow ranges for the ``solar'' and ``accelerator'' angles $\theta_{12}$ and $\theta_{23}$.
This translates into phenomenological implications for the expected neutrino and anti-neutrino appearance probabilities in neutrino oscillation experiments~\cite{Pasquini:2016kwk,Chatterjee:2017xkb}.
To illustrate the implications of our mixing patterns for future long baseline oscillation experiments
we present the resulting oscillation probabilities in Figs.~\ref{fig:pmueE} and \ref{fig:pmueL}. 

\begin{figure}[h!]
\begin{center}
\begin{tabular}{cc}
\includegraphics[width=0.8\linewidth]{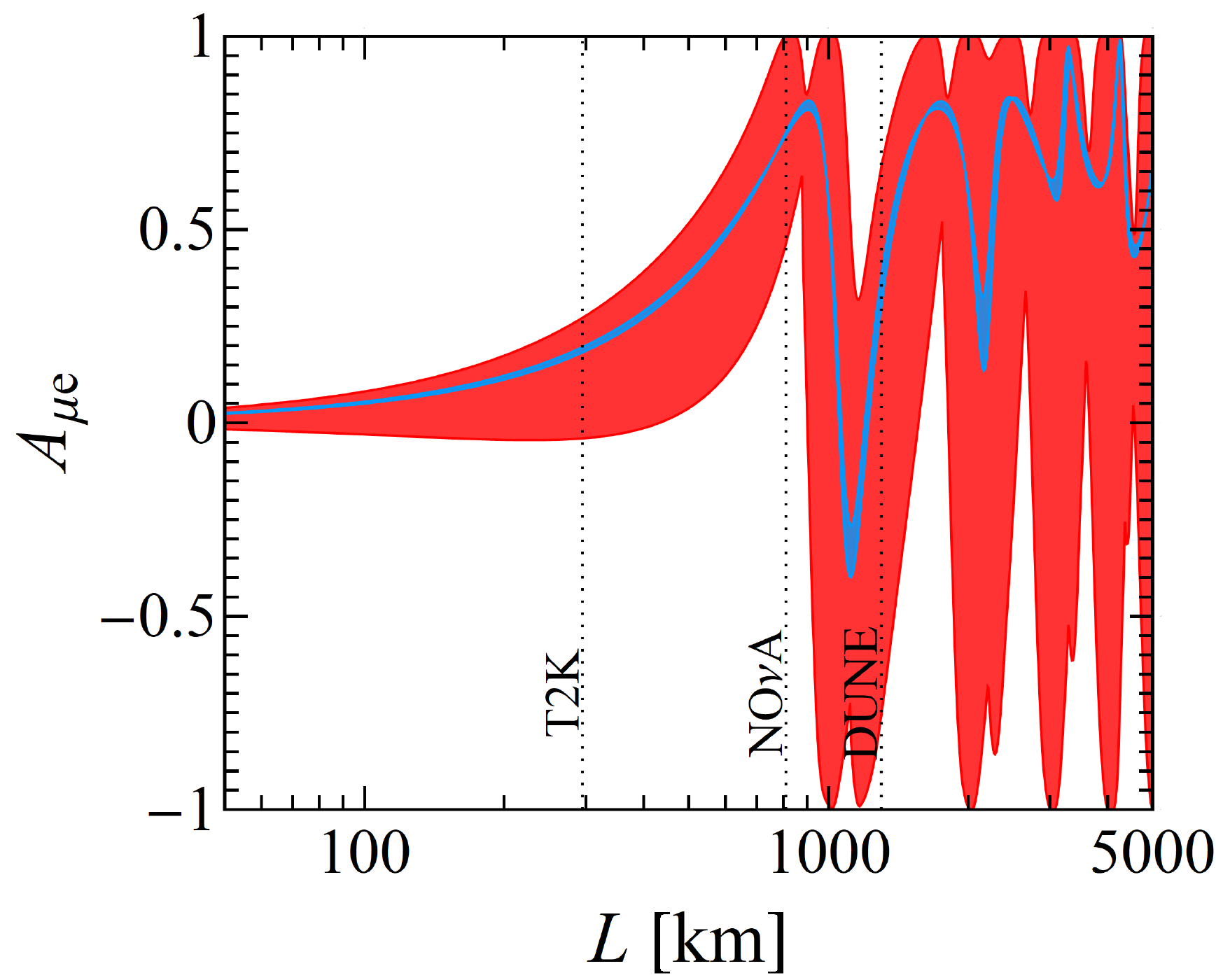}
\end{tabular}
\caption{\label{fig:amue}The CP asymmetry $A_{\mu e} = \frac{P(\nu_\mu \to \nu_e)-P(\bar{\nu}_\mu \to \bar{\nu}_e)}{P(\nu_\mu \to \nu_e) + P(\bar{\nu}_\mu \to \bar{\nu}_e)}$ versus distance taking the mixing angles and Dirac CP phase within the currently allowed 3$\sigma$ range. The broad band refer to a generic scenario, whereas the thin band refers to the bi-large pattern I prediction. }
\end{center}
\end{figure}

One sees that indeed the expected oscillation probabilities are tightly restricted, indicating that our bi-large mixing pattenrs should be testable at the upcoming long baseline oscillation experiments. In particular, the CP asymmetry, displayed in Fig.~\ref{fig:amue}, is very tightly predicted as compared to the generic three-neutrino oscillation scheme. This is seen by comparing the thin band (blue) with the broad band (red) in the figure.


\section{Conclusion}
\label{sec:conclusion}


In this letter we have proposed two bi-large-type lepton mixing schemes. They make definite assumptions on the two factors that comprise the lepton mixing matrix $U = U_l^\dagger U_\nu$, where $U = U_l$ comes from the charged sector while $U_\nu$ describes the neutrino diagonalization matrix.
We assume $U_\nu$ to be CP conserving and its three angles to be related with the Cabibbo angle in a simple way, given as $\sin\theta^\nu_{13}=\lambda$, $\sin\theta^\nu_{12}=2\lambda$ and $\sin\theta^\nu_{23}=1-\lambda$ (pattern I) or $3\lambda$ (pattern II), with the Dirac CP phase taken at the CP conserving value $\delta^\nu_{CP}=\pi$. CP violation arises only from the $U_l$ factor, assumed to have the CKM form, $U_l\simeq V_{\rm CKM}$, as expected in the simplest Grand Unified models. The Dirac CP phase is predicted as $\delta_{\rm CP}\sim 1.3\pi$, very close to its current best fit value. The mixing angles also depend on a single parameter $\phi$. The good measurement of the ``reactor'' angle leads to tight correlations that predict narrow ranges for the ``solar'' and ``accelerator'' angles $\theta_{12}$ and $\theta_{23}$ in good agreement with current oscillation data. The predictions should be testable at the upcoming long baseline oscillation experiments. Moreover, the structure of the two patterns is very simple, consistent with unification scenarios, and suggestive of novel model building approaches involving Abelian family symmetries~\cite{Ding:2012wh}.

\acknowledgments

PC is supported by National Natural Science Foundation of China under Grant No 11847240 and China Postdoctoral Science Foundation Grant No 2018M642700. GJD acknowledges the support of the National Natural Science Foundation of China under Grant No 11835013. RS and JV are supported by the Spanish grants SEV-2014-0398 and FPA2017-85216-P (AEI/FEDER, UE), PROMETEO/2018/165 (Generalitat Valenciana) and the Spanish Red Consolider MultiDark FPA2017-90566-REDC.

\providecommand{\href}[2]{#2}\begingroup\raggedright\endgroup

\end{document}